\PassOptionsToPackage{unicode}{hyperref}
\PassOptionsToPackage{hyphens}{url}
\documentclass[
]{article}
\usepackage{xcolor}
\usepackage{amsmath,amssymb}
\usepackage{iftex}
\ifPDFTeX
  \usepackage[T1]{fontenc}
  \usepackage[utf8]{inputenc}
  \usepackage{textcomp} %
\else %
  \usepackage{unicode-math} %
  \defaultfontfeatures{Scale=MatchLowercase}
  \defaultfontfeatures[\rmfamily]{Ligatures=TeX,Scale=1}
\fi
\usepackage{lmodern}
\ifPDFTeX\else
\fi
\IfFileExists{upquote.sty}{\usepackage{upquote}}{}
\IfFileExists{microtype.sty}{%
  \usepackage[]{microtype}
  \UseMicrotypeSet[protrusion]{basicmath} %
}{}
\makeatletter
\@ifundefined{KOMAClassName}{%
  \IfFileExists{parskip.sty}{%
    \usepackage{parskip}
  }{%
    \setlength{\parindent}{0pt}
    \setlength{\parskip}{6pt plus 2pt minus 1pt}}
}{%
  \KOMAoptions{parskip=half}}
\makeatother
\providecommand{\tightlist}{%
  \setlength{\itemsep}{0pt}\setlength{\parskip}{0pt}}
\usepackage{booktabs}
\usepackage{caption}
\usepackage{float}
\usepackage{tabularx}
\usepackage{tikz}
\usepackage{flowchart}
\usetikzlibrary{arrows.meta,fit,positioning}
\usepackage{bookmark}
\IfFileExists{xurl.sty}{\usepackage{xurl}}{} %
\makeatletter
\@ifundefined{xmpquote}{}{}
\makeatother
\hypersetup{
  pdftitle={FOI-O: A global ontology and verification framework for Freedom of Information process modelling},
  pdfauthor={Dylan A Mordaunt},
  hidelinks,
  pdfcreator={LaTeX via pandoc}}

\title{FOI-O: A global ontology and verification framework for Freedom
of Information process modelling}
\author{Dylan A Mordaunt\textsuperscript{1,2,3}}
\date{2026-07-16}

\begin{document}
\maketitle
\begin{abstract}
Public official-information request records contain process signals.
They can support research, workflow review, and analyst-led assessment.
Yet they also mix observed correspondence, platform states, inferred
events, and legal outcomes. FOI-O is a reusable process-modelling method
and verification infrastructure for Freedom of Information
administration. It is a global model that began with the New Zealand
Official Information Act and has since iterated through the Australian
Commonwealth and New South Wales settings. The NZ package remains the
mature reference implementation; the Australian work remains provisional
pending empirical evaluation and jurisdiction-specific legal validation.
FOI-O models request records, observed correspondence, controlled
vocabularies, provenance, review queues, release metadata, and bounded
analysis rules. Legally meaningful outcomes require certification by an
authorised decision-maker. Its typed operational and semantic contracts
are supported by deterministic examples, process models, fixture-only
process-mining exports, quality gates, and tests. This article describes
the motivation, architecture, ontology-development method, its versioned
extraction and review protocol, how related repositories share data and
evidence, and how the method may be adapted for Australia, validation
evidence, and implementation boundaries. The project is not legal
advice, is not an official government publication, and does not certify
release, refusal, redaction, charging, extension, transfer, complaint,
or publication outcomes.
\end{abstract}

\textsuperscript{1} Faculty of Health, Education and Psychology,
Victoria University of Wellington. \textsuperscript{2} College of
Medicine and Public Health, Flinders University. \textsuperscript{3}
Centre for Health Policy, The University of Melbourne.


\section{Introduction}\label{introduction}

Freedom of Information (FOI; see the
\hyperlink{tab-abbreviations}{abbreviations table}) administration has
many process steps. This is true across many legal systems. Yet the
process is often hard to compare across institutions and jurisdictions.
In this article, FOI refers to legal and administrative systems that
allow people to request information from public bodies. A single request
usually begins with intake: the request is received, registered,
acknowledged, and sometimes clarified or transferred to another body. It
then moves into decision-making, where deadlines, extensions, searches,
consultations, charges, redactions, releases, and refusals may need to
be recorded {[}\hyperlink{ref-1}{1}-\hyperlink{ref-3}{3}{]}. After the decision, the same request may still
matter for complaints, publication, disclosure logs, performance
statistics, and later reporting. Practitioners know these steps well.
They are not always recorded in a consistent machine-readable form.
Evidence may be spread across email, case-management tools, public
request platforms, disclosure logs, spreadsheets, attachments, and
statistical reports. As a result, the same process event can be visible
as a message in one system, a platform state in another, and a reporting
category in a third.

This makes FOI administration a useful but demanding target for process
modelling. Public request records show how requests move through real
administrative workflows. They can show when a request was made. They
can show whether an agency acknowledged it, sought clarification,
changed an apparent due date, released information, or faced a public
complaint. Public platforms such as FYI.org.nz (FYI; see the
\hyperlink{tab-abbreviations}{abbreviations table}) are especially
useful because they make parts of the request history visible outside
the agency. Even so, these records are platform-mediated evidence. They
are not agency systems of record. They can be incomplete, delayed,
duplicated, redacted, or unclear. A platform label may not match the
legal status of a request. A message timestamp may not be the statutory
date that matters. A visible attachment may not be the full agency
decision. Public archive and capture tools make parts of this record
available for research {[}\hyperlink{ref-4}{4},\hyperlink{ref-5}{5}{]}. Any reusable FOI data model therefore
needs to preserve what was observed while clearly marking what was
inferred {[}\hyperlink{ref-6}{6}-\hyperlink{ref-13}{13}{]}.

The problem is not only technical. FOI systems support democratic
accountability, public-sector learning, journalism, research, and
individual rights. When the process is hard to inspect, it is harder to
compare agency practice. It is also harder to explain delays, assess
reporting, or see where request handling could improve. A reusable FOI
process resource therefore needs to be clear to more than software
developers. It should describe the journey of a request in words that
policy analysts, lawyers, researchers, civic technologists, and public
officials can examine. It must also be precise enough for machines to
check examples, detect missing evidence, and support reproducible
analysis. Recent public-governance evidence supports this need. OECD
trust-survey work links trust in public institutions to transparency,
responsiveness, integrity, and evidence-informed decisions.
Open-government measurement frameworks also treat access to information
as a core part of accountable government {[}\hyperlink{ref-10}{10},\hyperlink{ref-13}{13},\hyperlink{ref-14}{14}{]}.

This also fits an older idea in democratic theory. Popper's open society
concept stresses public criticism, correction, and the ability to
challenge official power {[}\hyperlink{ref-8}{8}{]}. FOI gives that idea an administrative
form. It lets people ask what government has done, inspect the records
that support decisions, and test public claims against evidence.
Computation can help this work when it sorts, links, checks, and
explains records. It supports open government best when it widens
scrutiny without replacing human judgement or public accountability.

The same distinction matters for analyst-facing systems. Extraction,
retrieval, summary, and validation tools can help organise FOI material,
but a candidate signal may be mistaken for an official outcome. For
example, a text classifier might identify a message that resembles an
extension notice or a document that appears to contain released
information. That does not mean the software can certify whether the
extension, release, refusal, redaction, charge, transfer, or complaint
outcome was lawful or final. FOI-O addresses this risk by treating
public FOI workflow data as evidence for review, not as autonomous legal
decision-making. Its core distinction is between observed evidence,
candidate interpretation, and an outcome certified by an authorised
decision-maker. Analysts may use the software for routing, summary,
event extraction, evidence checks, and review preparation. Authorised
humans remain responsible for legally meaningful decisions.

The New Zealand Official Information Act (OIA; see the
\hyperlink{tab-abbreviations}{abbreviations table}) supplied the first
worked jurisdictional example. This origin is deliberate. New Zealand
provides a concrete legal and administrative setting that the author
understands, and FYI provides public request material that can test a
practical model {[}\hyperlink{ref-1}{1}-\hyperlink{ref-5}{5}{]}. FOI-O is now the global reusable method and
conceptual frame. FOI-O NZ remains its mature reference implementation.
Australian Commonwealth and New South Wales are the first two
provisional adaptations of the same core. They remain unvalidated until
each has its own sources, mappings, evaluation evidence, and approval
record.

FOI regimes differ in deadlines, exemptions, appeal paths, proactive
publication practice, reporting categories, and institutional culture.
They also share recurring process problems. Requests are received,
clarified, routed, answered, refused, extended, published, challenged,
and counted. A reusable ontology should make those common structures
visible while allowing local rules and terms to be added later. The
current contribution is a bounded, reproducible methods package. It is
not a live public-service system. It defines an auditable data model,
semantic layer, validation gates, local examples, repository
architecture, process architecture, data-model diagrams, release
metadata, and tests. These can be inspected without live credentials or
private request content. Together, they show the boundary between
repository-local proof and future external validation.

This article makes eight practical contributions. First, it gives
researchers and analysts a consistent way to record what happened in an
FOI request and what is only a suggested interpretation (schema-first
request and event contracts). Second, it supplies shared names and
machine-checkable rules so that different projects can describe the same
kinds of evidence (OWL, SKOS, RDF, and SHACL). Third, it makes clear
that software may organise evidence but cannot certify a legal or
administrative outcome (the certification boundary). Fourth, it provides
process diagrams and exchange files that people can inspect or move
between tools (BPMN and PNML). Fifth, it includes small, reproducible
examples showing how FOI events can be exchanged and checked by
process-mining software (XES and OCEL-style fixtures). Sixth, it records
a New Zealand study plan for annotation, while clearly stating that the
plan is not a result until independent review and adjudication evidence
exists. Seventh, it sets out a versioned procedure for extracting and
reviewing records, including fixed tool versions, evidence thresholds,
and approval gates. Eighth, it separates the shared model from local
legal and administrative settings (independently versioned core,
country, and subdivision profiles). The Australian Commonwealth and New
South Wales adaptations remain provisional pilots, not validated legal
profiles. Technical file paths and validation details are listed in the
supplement.

The article is organised as follows. The Methods section states the
design principles, repository architecture, ontology-development
protocol, process model, data model, and human-certification boundary.
The Results section describes what the current repository implements and
validates. The Discussion explains why this bounded architecture matters
for future comparative FOI research and accountable analyst-led public
administration.

\textbf{Version conventions.}

The version labels in this article refer to different things.
\textbf{FOI-O v0.8.1} is the latest published software release. The
\textbf{extraction and review protocol} is a separate research contract
describing how records may be prepared and reviewed; it is not a
software release. The \textbf{core, country, and subdivision profiles}
have their own compatibility versions, while source packs and legal
materials are versioned independently. A protocol milestone or planned
release identifier therefore does not mean that a new FOI-O release,
dataset, or validated legal profile exists.

\section{Methods}\label{methods}

\subsection{Design Principles}\label{design-principles}

FOI-O was developed around five design principles. The first two protect
the evidence record. Source evidence is kept before interpretation. This
means that observed labels, timestamps, correspondence records, platform
states, and attachment references are retained before they are mapped to
normalised process concepts. Observation is also kept separate from
certification. An extracted event may be useful for review, but it is
not treated as a final legal or administrative outcome unless a
human-authorised record supports that status.

The next two principles protect clarity and safety. Semantics remain
inspectable because schemas, controlled vocabularies, ontology files,
validation rules, mappings, examples, and tests are kept as reviewable
artefacts. They are not hidden inside a model or prompt. The system
fails closed around legal outcomes. When evidence is missing, unclear,
or generated by an automated component, the output is framed as a
candidate signal for review rather than a certified decision. The final
principle concerns proof. Local proof is preferred over aspirational
claims. Repository tests, examples, quality gates, and generated
metadata define what the package can currently show. Live services and
jurisdictional adoption remain external gates.

The design principles and implementation consequences are summarised in
\hyperlink{tab-design-principles}{Table 1}.

\begin{table}[H]
\small
\hypertarget{tab-design-principles}{}
\begin{center}\small\textbf{Table 1: Design principles used to develop the FOI-O methods package. Abbreviations: FOI, Freedom of Information.}\end{center}
\begin{tabularx}{\linewidth}{>{\raggedright\arraybackslash}p{0.34\linewidth}X}
\toprule
Principle & Implementation consequence \\
\midrule
Preserve source evidence & Keep observed labels, timestamps, and evidence before mapping to normalised states. \\
Separate observation from certification & Mark candidate events as reviewable signals, not final legal outcomes. \\
Keep semantics inspectable & Commit JSON Schema, SKOS, OWL, RDF, SHACL, mappings, and examples as reproducible artefacts. \\
Fail closed around legal outcomes & Reject autonomous certification of decision-like outcomes. \\
Prefer local proof & Use tests, examples, and validation commands to define what the repository can prove. \\
Track provenance & Retain source identifiers, content hashes, rights metadata, and transformation versions beside derived records. \\
\bottomrule
\end{tabularx}
\end{table}

\subsection{Repository Architecture}\label{repository-architecture}

The architecture follows a source, archive, semantic, analysis, and
evaluation layering pattern. Source request records and archive
manifests are preserved upstream. FOI-O maps those records into request
profiles, event streams, and controlled vocabularies. It also maps them
into Resource Description Framework (RDF; see the
\hyperlink{tab-abbreviations}{abbreviations table}) and Shapes
Constraint Language (SHACL; see the
\hyperlink{tab-abbreviations}{abbreviations table}) artefacts, plus
bounded analytical resources {[}\hyperlink{ref-15}{15}-\hyperlink{ref-20}{20},\hyperlink{ref-21}{21},\hyperlink{ref-22}{22}{]}.

The wider programme {[}\hyperlink{ref-23}{23}{]} deliberately separates capture, archival
fidelity, document processing, candidate extraction, ontology contracts,
deterministic rules, and programme conformance. \texttt{fyi-cli}
captures FYI/Alaveteli-compatible source and delta inputs {[}\hyperlink{ref-4}{4}{]}.
\texttt{fyi-archive} preserves manifests and provenance, packages
datasets, and publishes versioned outputs to Hugging Face and
preservation services {[}\hyperlink{ref-5}{5},\hyperlink{ref-24}{24}{]}. \texttt{foi-process} provides the
integration spine for document evidence and optical character
recognition (OCR) {[}\hyperlink{ref-25}{25}{]}. \texttt{nlp-policy-nz} evaluates
review-bounded extraction adapters {[}\hyperlink{ref-26}{26}{]}. The \texttt{legislation}
repository supplies versioned statutory source packs {[}\hyperlink{ref-27}{27}{]}, while
\texttt{rulespec-nz} supplies deterministic New Zealand rule
specifications {[}\hyperlink{ref-28}{28}{]}. \texttt{rac-conformance} synchronises
cross-repository conformance evidence {[}\hyperlink{ref-29}{29}{]}. FOI-O consumes pinned,
provenance-bearing inputs from these surfaces; it does not collapse them
into one application or treat downstream model output as a certified
legal record {[}\hyperlink{ref-23}{23}{]}.

\hyperlink{fig-repository-architecture}{Figure 2} shows the
repository-level architecture, while
\hyperlink{fig-process-architecture}{Figure 3} shows the process flow
through validation and the human-certification boundary.

\begin{figure}[H]
\centering
\resizebox{\linewidth}{!}{%
\begin{tikzpicture}[
  node distance=1.05cm and 1.15cm,
  every node/.style={font=\small, align=center},
  flow/.style={-Latex, line width=0.7pt},
  support/.style={-Latex, dashed, line width=0.65pt, draw=gray!70},
  asset/.style={storage, draw, minimum width=3.2cm, minimum height=1.0cm, fill=gray!10},
  contract/.style={process, draw, minimum width=3.2cm, minimum height=1.0cm, fill=blue!6},
  semantic/.style={predproc, draw, minimum width=3.3cm, minimum height=1.0cm, fill=purple!8},
  runtime/.style={process, draw, minimum width=3.2cm, minimum height=1.0cm, fill=orange!10},
  output/.style={terminal, draw, minimum width=3.25cm, minimum height=1.0cm, fill=green!8},
  qa/.style={process, draw, minimum width=3.45cm, minimum height=0.9cm, fill=yellow!12}
]
\node[asset] (docs) {Jurisdiction and\\method documents};
\node[asset, below=of docs] (examples) {Fixtures, examples,\\and mappings};
\node[contract, right=of docs] (schemas) {Schemas and\\data models};
\node[contract, right=of examples] (events) {Request profiles\\and event contracts};
\node[semantic, right=of schemas] (ontology) {Ontology and\\vocabularies};
\node[semantic, right=of events] (shacl) {RDF export and\\SHACL constraints};
\node[runtime, right=of ontology] (cli) {Command-line\\workflows};
\node[runtime, right=of shacl] (quality) {Validation and\\quality gates};
\node[output, right=of cli] (publication) {Release\\metadata};
\node[output, right=of quality] (analyst) {Read-only analyst\\workspaces};
\node[qa, below=1.05cm of quality] (tests) {Tests, examples, and release checks};
\node[contract, below=1.05cm of events] (provenance) {Provenance, hashes,\\and rights metadata};

\draw[flow] (docs) -- (schemas);
\draw[flow] (examples) -- (events);
\draw[flow] (schemas) -- (ontology);
\draw[flow] (events) -- (shacl);
\draw[flow] (ontology) -- (cli);
\draw[flow] (shacl) -- (quality);
\draw[flow] (cli) -- (publication);
\draw[flow] (quality) -- (analyst);
\draw[flow] (schemas) -- (events);
\draw[flow] (ontology) -- (shacl);
\draw[flow] (cli) -- (quality);
\draw[flow] (events) -- (provenance);
\draw[support] (provenance) -| (quality.south);
\draw[support] (tests) -| (schemas.south);
\draw[support] (tests) -- (quality);
\draw[support] (tests) -| (publication.south);
\end{tikzpicture}%
}
\hypertarget{fig-repository-architecture}{}
\begin{center}\small\textbf{Figure 2: FOI-O repository architecture. The repository contains reviewable documents and fixtures, machine-readable contracts, provenance and rights metadata, semantic assets, runtime workflows, release metadata, analyst workspaces, and tests that bind the layers together. Abbreviations: FOI-O, Freedom of Information Ontology; RDF, Resource Description Framework; SHACL, Shapes Constraint Language.}\end{center}
\end{figure}

\begin{figure}[H]
\centering
\resizebox{\linewidth}{!}{%
\begin{tikzpicture}[
  node distance=1.35cm and 1.35cm,
  every node/.style={font=\small, align=center},
  flow/.style={-Latex, line width=0.7pt},
  feedback/.style={-Latex, dashed, line width=0.65pt, draw=gray!70},
  store/.style={storage, draw, minimum width=3.0cm, minimum height=1.0cm, fill=gray!10},
  proc/.style={process, draw, minimum width=3.0cm, minimum height=1.0cm, fill=blue!6},
  check/.style={decision, draw, aspect=2, minimum width=2.6cm, minimum height=1.2cm, fill=orange!10},
  guard/.style={terminal, draw, minimum width=3.4cm, minimum height=1.0cm, fill=green!8},
  boundary/.style={draw, dashed, rounded corners=2pt, inner sep=6pt}
]
\node[store] (source) {Public request\\records};
\node[store, below=of source] (archive) {Archive\\manifests};
\node[proc, right=of source] (profile) {Request\\profiles};
\node[proc, right=of archive] (events) {Candidate\\process events};
\node[check, right=1.45cm of profile] (validation) {Schema and\\model checks};
\node[proc, right=1.45cm of validation] (semantic) {Vocabularies\\and SHACL};
\node[guard, right=1.45cm of semantic] (analyst) {Read-only\\analyst review pack};
\node[guard, below=1.25cm of analyst] (human) {Authorised\\process decision};

\draw[flow] (source) -- (profile);
\draw[flow] (archive) -- (events);
\draw[flow] (profile) -- (validation);
\draw[flow] (events.east) -- ++(0.72cm,0) |- (validation.south);
\draw[flow] (validation) -- node[above, midway]{validated} (semantic);
\draw[flow] (semantic) -- (analyst);
\draw[flow] (analyst) -- node[right]{review only} (human);
\draw[feedback] (human.south west) -- ++(0,-0.55cm) -| node[below,pos=0.25, text=gray!70]{certification evidence} (events.south);

\node[boundary, fit=(analyst) (human), label={[font=\small]above:human certification boundary}] {};
\end{tikzpicture}%
}
\hypertarget{fig-process-architecture}{}
\begin{center}\small\textbf{Figure 3: FOI-O process architecture. Public request platforms and archive manifests are transformed into request profiles and candidate process events, checked through schema and model validation, aligned with semantic vocabularies and constraints, and packaged for analyst review before an authorised process decision. Abbreviations: FOI-O, Freedom of Information Ontology; SHACL, Shapes Constraint Language.}\end{center}
\end{figure}

The Python control plane owns schema validation, FYI manifest
normalisation, event extraction, quality gates, reporting profiles, RDF
export, SHACL validation, release metadata, and command-line workflows.
Optional surfaces add runtime capability when installed. They do not
define the core proof. Deterministic Python paths and fixtures remain
the reproducibility base. FastMCP can expose read-only tools in an
analyst workspace. pySHACL can validate RDF graphs against SHACL
constraints. LanceDB is an embedded vector database. It could support
local semantic retrieval over request text, evidence chunks, and
ontology terms. Mojo and Modular MAX are being explored as local
high-performance inference tools. They could later support bounded
extraction or embedding workflows. They are not required for the current
package. The present evidence depends on portable schemas, examples,
semantic assets, and Python tests, not on specialised runtimes,
hardware, or model-serving installs.

Data provenance is carried through each step. The source manifest,
archive or capture identifier, content hash, and rights restriction are
retained before a record is normalised. Deterministic transformations
then create request profiles, observed messages, and candidate events;
each derived item points back to its source evidence and records the
code, profile, and transformation versions used. Validation checks the
derived data without turning a candidate event into a certified outcome.
The repository locations and reproducible commands are detailed in the
supplement.

\subsection{Empirical Extraction and Jurisdiction
Profiles}\label{empirical-extraction-and-jurisdiction-profiles}

The versioned protocol extends rather than replaces the earlier method.
The earlier method established source provenance, epistemic status,
evidence references, validation, and the human-certification boundary.
The newer protocol adds capability declarations and promotion evidence.
An adapter can be promoted only when its source, profile, model, and
transformation versions are immutably identified; its evaluation data
have a recorded rights basis; independent annotation and adjudication
are documented; per-capability metrics meet declared thresholds; and an
authorised reviewer explicitly approves the promotion. Missing evidence
fails closed. A syntactically valid adapter is therefore not necessarily
an empirically supported or legally approved profile.

FOI-O uses a versioned ontology family rather than long-lived
jurisdiction Git branches. \texttt{foi-o} identifies the
jurisdiction-neutral core. Country profiles, such as \texttt{foi-o-nz}
and the planned \texttt{foi-o-au}, declare compatible core-version
ranges. Subdivision profiles, such as \texttt{foi-o-au-nsw}, also
declare a compatible parent-country profile. The existing Python
distribution remains the implemented NZ package while these contracts
mature. Australian Commonwealth and New South Wales are the first
provisional pilots. They are not promoted legal profiles, and the
remaining Australian states and territories remain disabled until each
has jurisdiction-specific legislation, examples, annotation, evaluation,
compatibility evidence, and jurisdiction-specific legal validation
{[}\hyperlink{ref-26}{26},\hyperlink{ref-27}{27}{]}.

\subsection{Ontology Development
Protocol}\label{ontology-development-protocol}

The ontology-development protocol uses repository evidence as the source
of truth. The first step is scoping. FOI-O identifies process concepts
from FOI request workflows, FYI/Alaveteli source states,
statutory-process concepts, release metadata, and reporting needs. New
Zealand OIA material is used as the first worked example. This produces
a practical concept inventory before the project attempts richer formal
modelling.

The second step is operational encoding. The project encodes the minimum
contracts as JavaScript Object Notation (JSON; see the
\hyperlink{tab-abbreviations}{abbreviations table}) Schemas and Pydantic
models before adding semantic alignments. This makes examples and
command outputs testable early. Controlled vocabularies are then defined
for request states, event types, assertion status, and review boundaries
using the Simple Knowledge Organization System (SKOS; see the
\hyperlink{tab-abbreviations}{abbreviations table}).

The third step is semantic alignment. Event and evidence concepts are
aligned with the Provenance Ontology (PROV-O; see the
\hyperlink{tab-abbreviations}{abbreviations table}). This keeps
transformations and sources traceable. Dataset and publication concepts
are aligned with the Data Catalog Vocabulary (DCAT; see the
\hyperlink{tab-abbreviations}{abbreviations table}). Rights and policy
concepts use the Open Digital Rights Language (ODRL; see the
\hyperlink{tab-abbreviations}{abbreviations table}), SKOS, and
legal-document references where appropriate {[}\hyperlink{ref-15}{15}-\hyperlink{ref-20}{20}{]}. Safety and
consistency constraints are expressed in SHACL. The resulting examples
and code paths are validated through tests, release-readiness checks,
and machine-readable release metadata.

The core event contract defines observed and candidate process events.
Semantic constraints define how machine-readable process statements are
checked before they are used in reports, review packs, or analytical
workflows.

\subsection{Ontology, Process Model, and Interchange
Artefacts}\label{ontology-process-model-and-interchange-artefacts}

FOI-O separates the semantic model from the workflow model. The
ontology, controlled vocabularies, RDF exports, and SHACL shapes form
the semantic layer. They define terms, relationships, provenance
expectations, and safety constraints that can be inspected independently
of any one runtime.

The process-model layer describes how requests can move through
lifecycle states and event families. It includes a state machine, BPMN
process-model artefacts, and PNML Petri-net artefacts. These models are
review and interchange artefacts. They are not executable legal
workflows, and they do not certify access, refusal, release, redaction,
charging, extension, transfer, or complaint outcomes.

The process-mining interchange layer is narrower still. It contains a
committed fixture event log, XES and OCEL-style exports, and a
fixture-only conformance report for one release path {[}\hyperlink{ref-21}{21},\hyperlink{ref-22}{22}{]}. These
artefacts demonstrate import, interchange, and boundary-preserving
conformance checks on a deterministic fixture. They do not show
live-corpus process discovery, agency bottlenecks, cycle-time
distributions, or real-world process conformance.

\subsection{Data Model}\label{data-model}

The model is organised in three groups. The request and event core
describes request profiles, process events, evidence references,
assertion status, provenance, generator metadata, and
human-certification metadata. This is the part of the model that records
what happened, where the evidence came from, and how strongly a process
statement can be asserted.

The review and governance layer describes recorded analysis actions,
review tasks, ledgers, chunks, and risk assessments. These records show
how candidate outputs were generated and reviewed, without turning them
into legal determinations. The reporting and release layer describes
reporting metrics and release metadata. Candidate process events can
support triage and review, but they are not promoted to certified
outcomes unless an authorised human record supplies the certification
evidence.

\hyperlink{fig-data-model}{Figure 4} shows the main data-model
relationships, and \hyperlink{tab-data-model-surfaces}{Table 5}
summarises the corresponding model surfaces.

\begin{figure}[H]
\centering
\resizebox{\linewidth}{!}{%
\begin{tikzpicture}[
  node distance=1.2cm and 1.55cm,
  every node/.style={font=\small, align=center},
  relation/.style={-Latex, line width=0.7pt},
  optional/.style={-Latex, dashed, line width=0.65pt, draw=gray!70},
  entity/.style={process, draw, minimum width=3.2cm, minimum height=1.0cm, fill=blue!6},
  evidence/.style={storage, draw, minimum width=3.2cm, minimum height=1.0cm, fill=gray!10},
  semanticnode/.style={predproc, draw, minimum width=3.4cm, minimum height=1.0cm, fill=purple!8},
  gate/.style={terminal, draw, minimum width=3.3cm, minimum height=1.0cm, fill=green!8},
  metric/.style={process, draw, minimum width=3.2cm, minimum height=1.0cm, fill=orange!10}
]
\node[entity] (request) {Request profile};
\node[entity, right=of request] (event) {Process event};
\node[evidence, below=of event] (evidence) {Evidence reference};
\node[semanticnode, right=1.65cm of event] (assertion) {Assertion status\\and provenance};
\node[gate, right=1.75cm of assertion] (human) {Human certification\\metadata};
\node[entity, above=of assertion] (analyst) {Analysis action\\record};
\node[metric, below=of assertion] (review) {Review task\\or ledger entry};
\node[metric, right=of review] (reporting) {Reporting metric\\or release metadata};

\draw[relation] (request) -- (event);
\draw[relation] (event) -- node[right]{cites} (evidence);
\draw[relation] (event) -- (assertion);
\draw[optional] (analyst) -- node[right]{may generate} (assertion);
\draw[optional] (assertion) -- (human);
\draw[relation] (assertion) -- (review);
\draw[relation] (review) -- (reporting);
\draw[optional] (human) |- (reporting);
\end{tikzpicture}%
}
\hypertarget{fig-data-model}{}
\begin{center}\small\textbf{Figure 4: FOI-O data model. Request profiles organise observed and candidate process events. Events cite evidence, carry assertion status and provenance, and may be generated or reviewed through recorded analysis actions. Certified status requires an authorised certification record; review tasks, ledgers, reporting metrics, and release metadata remain downstream artefacts. Abbreviations: FOI-O, Freedom of Information Ontology.}\end{center}
\end{figure}

\begin{table}[H]
\small
\hypertarget{tab-data-model-surfaces}{}
\begin{center}\small\textbf{Table 5: Main data-model surfaces in the FOI-O implementation. Abbreviations: JSON-LD, JavaScript Object Notation for Linked Data; RDF, Resource Description Framework; SHACL, Shapes Constraint Language.}\end{center}
\begin{tabularx}{\linewidth}{>{\raggedright\arraybackslash}p{0.32\linewidth}X}
\toprule
Surface & Current role \\
\midrule
Request profile schema & Request metadata, source state, identifiers, and JSON-LD context. \\
Core event schema & Observed or candidate process events with evidence and assertion status. \\
SHACL shapes & Semantic validation and safety constraints over RDF exports. \\
Review action schema & Preparatory analytical outputs bounded away from legal certification. \\
Process-model artefacts & State machine, BPMN, and PNML review/interchange models for workflow structure. \\
Process-mining fixtures & Deterministic event-log, XES, OCEL-style, and fixture-conformance artefacts. \\
Empirical study plan & Planned New Zealand annotation tasks, not an independently reviewed gold standard. \\
Release metadata & Evidence, rights notices, and external publication gates. \\
\bottomrule
\end{tabularx}
\end{table}

\subsection{Human Certification
Boundary}\label{human-certification-boundary}

The certification boundary is deliberately repeated. It appears in
schemas, model logic, quality gates, review policies, Model Context
Protocol (MCP; see the
\hyperlink{tab-abbreviations}{abbreviations table}) and tool
descriptors, SHACL shapes, examples, tests, and release metadata.
Analysts may map observed states, propose candidate events, assemble
review packs, compute indicative clocks, and check evidence
completeness. They must not certify legal outcomes.

This boundary is a scientific and governance constraint, not a cosmetic
warning. It prevents evaluation metrics, local extraction, retrieval, or
analytical contracts from being represented as legal determinations.

\section{Results}\label{results}

\subsection{Implemented Repository
Surfaces}\label{implemented-repository-surfaces}

The current repository includes implemented surfaces in seven groups.
The contract layer covers JSON Schema examples, Pydantic models, state
mapping, and manifest normalisation. The analysis layer covers event
analytics, quality gates, RDF export, and reporting profiles. The
release layer covers release metadata and reproducibility manifests. The
analyst-support layer covers local retrieval, redaction candidates,
review context packs, stream diffs, and read-only tool descriptors. The
process-model layer covers state-machine, BPMN, PNML, XES, OCEL-style,
and fixture-conformance artefacts. The Mojo, Modular MAX, and LanceDB
paths remain experimental and optional {[}\hyperlink{ref-23}{23}{]}. The empirical-contract
layer covers capability declarations, immutable dependency requirements,
evidence-count thresholds, and provisional promotion states. The
jurisdiction layer covers versioned profile manifests, parent/core
compatibility rules, and fail-closed Australian pilot declarations.

The first result is a set of machine-readable contracts. They make the
intended FOI-O data surfaces explicit. Request profiles define
request-level metadata, source state, identifiers, and linked-data
context. Core event contracts define observed or candidate process
events. These events include evidence references, assertion status,
provenance, generator metadata, and human-certification metadata. Review
action contracts define preparatory outputs that can help with review
without representing a final legal outcome. Deterministic examples show
how the model should behave on concrete records. This matters because
the package does not ask users to trust an invisible extraction prompt
or a private database. It exposes the shape of the data and the
validation expectations.

The second result is a semantic layer. It connects operational data
contracts to inspectable vocabularies and ontology artefacts. Controlled
vocabularies make state, event, assertion, and review-boundary terms
easier to inspect. RDF export and SHACL constraints provide a path from
local examples to semantic validation. Not every future user needs to
run a full semantic-web stack. The point is that the project can state,
in a machine-readable way, which relationships and safety constraints
should hold. For example, a candidate event can carry evidence and
provenance. Certified status requires a different kind of
human-authorised record. That distinction is shown in the model,
figures, and quality gates {[}\hyperlink{ref-15}{15}-\hyperlink{ref-20}{20}{]}.

The third result is a validation and quality-gate layer. The repository
checks examples, Python behavior, semantic alignment, release metadata,
release evidence, and formatting or packaging constraints. These checks
provide repository-local proof that the current package is internally
consistent. They do not prove that every public request platform has
been ingested. They do not prove that all New Zealand agencies are
represented. They also do not prove that the model has been validated
across other jurisdictions. Instead, they define a smaller and more
defensible result: the implemented contracts, examples, diagrams, and
validation artefacts can be rebuilt and checked locally.

The fourth result is a release and reuse surface. Release metadata,
examples, figures, glossary terms, and abbreviations make the package
easier to inspect. This matters because an ontology or validation stack
is hard to reuse if the evidence, caveats, rights notices, and human
gates are not packaged with it. The current package records what is
implemented, what is experimental, what requires external approval, and
what should not be treated as legal or operational certification.

The fifth result is a process-model and process-mining fixture surface.
The repository contains workflow process models in BPMN and PNML,
generated state-machine exports, a fixture event log, XES and OCEL-style
exports, and a fixture-only conformance report. These artefacts make the
process layer easier to inspect and easier to exchange with
process-analysis tools. Their claim is intentionally narrow: they show
that the committed fixture path preserves the human-certification
boundary and can be exported for review. They do not prove live-source
process discovery, process performance, or agency conformance.

The sixth result is a versioned extraction and review protocol.
Capability declarations make each adapter's claimed task surface
inspectable instead of inferring it from a model name or prompt.
Contract validation can reject absent provenance, insufficient evidence,
incompatible versions, or an attempted promotion that lacks human
approval. This is a governance result, not an accuracy claim: the
current archive, Commonwealth, and New South Wales adaptations remain
provisional until real heldout evaluation evidence satisfies those
gates.

The seventh result is a jurisdiction-versioning contract. The repository
can represent a shared core, independently released country profiles,
and independently released subdivision profiles without allowing one
jurisdiction to silently redefine a core concept. Compatibility ranges
and explicit migrations make legal and semantic change visible. The
contract does not claim that Australian law has already been validated;
it records exactly what must be supplied before such a claim is
possible.

The optional runtime surfaces are deliberately bounded. LanceDB is
relevant because FOI-O may need to retrieve similar requests, evidence
snippets, ontology terms, and review examples. It should be able to do
this without sending sensitive material to external services. Mojo and
Modular MAX are relevant because larger extraction and embedding
workflows may benefit from fast local inference once the data contracts
and safety boundaries are stable. The potential benefits are faster
local processing, less dependence on hosted providers, more reproducible
retrieval tests, and a clearer path to privacy-preserving analysis.
These tools are not currently used as the core proof. They add optional
dependencies, platform constraints, and model-selection questions. Those
issues are not needed to validate the ontology, schemas, examples, and
human-certification boundary. The core reproducibility claim therefore
rests on deterministic Python paths, examples, schemas, semantic assets,
and tests. This makes the current package useful as a baseline for later
corpus intake, planned annotation task-set evaluation, jurisdictional
extension, or analyst-led review. It also avoids unsupported claims
about live deployment readiness {[}\hyperlink{ref-23}{23}{]}.

Taken together, these results show a working methods package rather than
a finished operational FOI platform. The implemented surfaces
demonstrate that the main concepts can be represented, validated,
documented, and packaged in a repeatable form. The remaining work is to
test those surfaces against larger corpora, future jurisdictional
profiles, and independently reviewed evaluation sets.

\hyperlink{tab-evidence-surfaces}{Table 6} summarises the implemented
evidence surfaces and the validation evidence currently available in the
repository.

\begin{table}[htbp]
\small
\hypertarget{tab-evidence-surfaces}{}
\begin{center}\small\textbf{Table 6: Implemented evidence surfaces and validation evidence. Abbreviations: RDF, Resource Description Framework; SHACL, Shapes Constraint Language.}\end{center}
\begin{tabularx}{\linewidth}{>{\raggedright\arraybackslash}p{0.26\linewidth}>{\raggedright\arraybackslash}p{0.34\linewidth}X}
\toprule
Evidence surface & Evidence type & Validation evidence \\
\midrule
Schemas and examples & Machine-readable contracts and deterministic examples & Example validation suite \\
Core Python behavior & Executable implementation and regression tests & Unit-test suite \\
Release readiness & Release evidence and quality gates & Lint, format, and test checks \\
Release metadata & Versioned release and reuse records & Metadata tests \\
Semantic alignment & Ontology, vocabulary, and semantic constraints & SHACL safety tests \\
Process models & State-machine, BPMN, and PNML workflow artefacts & Process-model parsing and conformance tests \\
Process-mining fixtures & Fixture event log, XES, OCEL-style export, and fixture conformance & Fixture-only import and conformance tests \\
Empirical task-set plan & Planned New Zealand annotation tasks & Schema validation and boundary tests \\
Versioned extraction protocol & Capability and promotion-evidence declarations & Fail-closed contract and profile tests \\
Jurisdiction profiles & Core, country, and subdivision compatibility manifests & Version, parent, and promotion-boundary tests \\
\bottomrule
\end{tabularx}
\end{table}

\begin{table}[htbp]
\small
\hypertarget{tab-evidence-boundaries}{}
\begin{center}\small\textbf{Table 7: What the current package proves and does not prove. Abbreviations: FOI-O, Freedom of Information Ontology.}\end{center}
\begin{tabularx}{\linewidth}{>{\raggedright\arraybackslash}p{0.28\linewidth}>{\raggedright\arraybackslash}p{0.33\linewidth}X}
\toprule
Surface & What this proves & What it does not prove \\
\midrule
FOI-O NZ profile & Local schemas, examples, semantic assets, and tests are internally consistent. & Non-NZ validation or official adoption. \\
Australian adaptations & Commonwealth and New South Wales provisional contracts can be represented and checked. & Legal approval, empirical promotion, or coverage of other states and territories. \\
Process models & Workflow structures can be represented as review/interchange artefacts. & Executable legal workflow or certified decision-making. \\
Process-mining fixtures & Deterministic fixture events can be exported and checked for one release path. & Live-corpus conformance, bottlenecks, cycle times, or agency performance. \\
Empirical study plan & Planned annotation tasks and external gates are explicit. & Independently reviewed gold-standard evidence. \\
Release package & Repo-local publication and reuse evidence can be validated. & Journal, registry, arXiv, or government approval. \\
\bottomrule
\end{tabularx}
\end{table}

\clearpage

\section{Discussion}\label{discussion}

The evidence boundaries are summarised in
\hyperlink{tab-evidence-boundaries}{Table 7}.

FOI-O is designed to describe and check the process evidence around a
FOI request. It asks what was observed, where the evidence came from,
which candidate event it might support, and whether an authorised
reviewer checked it. It does not decide whether an agency acted
lawfully, whether a refusal ground was valid, or whether a release,
redaction, charge, transfer, extension, or complaint outcome was legally
final. This distinction matters for analyst-facing systems in any
jurisdiction. The same event stream that helps a reviewer find missing
evidence could be misused. This can happen if a model treats candidate
events as certified statutory outcomes. The repository therefore keeps
assertion status, provenance, review status, and certification metadata
close to the data model.

The value of this separation is broader than New Zealand. FOI systems
differ, but most regimes have some version of observable activity,
process interpretation, and authorised decision-making. A public
platform may show that a response was sent. A model may infer that the
response looks like a release or refusal. An authorised reviewer may
certify the administrative or legal meaning of that response. Treating
those three layers as the same thing may seem easier at first. It makes
later comparison weaker and less trustworthy. FOI-O instead makes the
layers explicit. That creates more modelling work, but it protects later
analysis from claiming more than the evidence can support.

This structure also helps future comparison. Public-information regimes
use different names, clocks, exemptions, appeal paths, publication
duties, and reporting rules. A reusable ontology cannot assume that
every future jurisdiction shares New Zealand terms or deadlines. It can,
however, provide a common pattern for the evidence trail. That trail
includes requests, observed correspondence, candidate events,
provenance, assertion status, review tasks, release metadata, and
human-certification boundaries. Future jurisdiction-specific profiles
can then add local vocabulary, calendar rules, statutory references,
reporting categories, and quality gates. In this sense, New Zealand is a
bootstrap case. It gives the project a concrete starting point. The
purpose is to make later comparative work easier {[}\hyperlink{ref-1}{1}-\hyperlink{ref-14}{14}{]}.

For public agencies and researchers, the approach could support several
uses. As an explanatory resource, it could show how FOI requests move
through public systems and where evidence is usually created. As a
process-analytics resource, it could help identify missing evidence,
unclear status changes, or public records that do not support strong
claims. As a comparative resource, it could help map reporting
categories across jurisdictions without forcing them into one legal
vocabulary. As an analyst resource, it could support tools that prepare
review packs, flag incomplete evidence, or summarise event streams. It
would still make clear that final legal or administrative judgement
remains with authorised people.

The schema-first approach has practical strengths. JSON Schema and
Pydantic models make the operational contract easy to test before richer
semantic alignment is attempted. They give quick feedback when examples,
command outputs, or release metadata drift from the expected structure.
SKOS vocabularies make state and event terms easier to inspect. Users do
not need to read application code to review them. RDF and SHACL add a
path to semantic validation for users who need it. Basic checks do not
have to depend on a heavier semantic-web environment. This layered
approach is useful for an early reusable infrastructure project because
simple local checks and richer ontology work can coexist {[}\hyperlink{ref-15}{15}-\hyperlink{ref-20}{20}{]}.

There are tradeoffs. A schema-first package can be easier to test and
adopt, but it may at first miss some legal and administrative detail. A
semantic-web-first package can express more formal relationships. It may
also be harder for agencies, researchers, or civic technologists to run
and inspect. FOI-O takes a middle path. Operational contracts define the
minimum reproducible data surfaces. Vocabularies, ontology files, and
SHACL constraints provide semantic alignment. This is not a claim that
the current model is complete. It is a claim that gaps can be found,
reviewed, and extended.

The same tradeoff appears in reproducibility. Local checks are valuable.
They let readers inspect the figures and tables, validate examples, and
check the human-boundary claims without live credentials. Local
reproducibility is not the same as live validation. The current
repository can prove local contracts, examples, and deterministic
transformations. It does not yet prove live archive intake or
independently reviewed gold-standard performance. The FOI-O v0.8.1
software release is preserved in Zenodo with version DOI
\href{https://doi.org/10.5281/zenodo.21360138}{10.5281/zenodo.21360138}
and concept DOI
\href{https://doi.org/10.5281/zenodo.21360137}{10.5281/zenodo.21360137}
{[}\hyperlink{ref-30}{30}{]}. That preservation evidence does not prove agency-internal
reporting completeness or transferability to every FOI regime. Those
claims should remain external gates until they are supported by
live-source evidence, independent review, jurisdiction-specific mapping,
and separate validation.

The strongest contribution of FOI-O is methodological. It shows how a
public-information process ontology can be built around evidence
preservation, bounded inference, human certification, semantic
inspection, and reproducible validation {[}\hyperlink{ref-21}{21}-\hyperlink{ref-23}{23}{]}. The contribution is
modest in operational scope, but it matters for future work. FOI systems
are increasingly analysed with automated tools. The infrastructure
around those tools must make clear which statements are observed,
inferred, validated, or certified by humans. FOI-O provides one concrete
way to encode that boundary and extend it beyond the first New Zealand
example.

Future work should proceed in evidence-led stages. The next step is not
to claim universal coverage. It is to re-extract pinned archive releases
through the versioned protocol and publish dataset outputs with complete
lineage. Gold-set review should only follow completed annotation tasks,
recorded reviewer process, adjudication, and provenance. Later
jurisdiction-specific profiles should add local law, language,
calendars, reporting rules, and institutional practice without changing
the shared evidence model. Comparative reporting examples should show
which metrics are truly comparable and which remain tied to one
jurisdiction. Each extension should preserve the same boundary between
observed record, candidate inference, validation result, and
human-certified outcome. That discipline is what makes later comparison
possible.

\section{Limitations}\label{limitations}

FOI-O is not legal advice, is not an official government publication,
and is not an official reporting system for New Zealand or any other
jurisdiction. It does not retrieve live source systems by default,
republish source FYI/archive payloads, replace agency records, decide
statutory interpretation, or certify FOI outcomes. It should be treated
as a research and validation artefact until live-source ingestion,
jurisdiction-specific mappings, and operational use are separately
reviewed.

The main threats to validity are scope and evidence limits. FOI-O's
global scope is architectural and methodological, not a claim of
universal legal validation. FOI-O NZ is the mature reference
implementation, while Australian Commonwealth and New South Wales
adaptations remain provisional iterations. The repository evidence is
local and fixture-heavy. It does not yet prove live archive intake,
representative agency coverage, independently reviewed gold-standard
labels, real-world process conformance, bottleneck frequencies, agency
cycle times, or corpus-level process-mining results. Process-mining
artefacts are included to show deterministic interchange and
fixture-path conformance only. The planned empirical task sets remain
annotation tasks until source snapshots, review instructions, human
labels, adjudication, and any agreement metrics are recorded.

\section{Conclusion}\label{conclusion}

FOI-O provides a global ontology and validation framework for FOI
analysis. It began with New Zealand and has iterated through Australian
jurisdictions without treating jurisdictional law as globally
interchangeable. Its strongest contribution is evidence discipline:
schemas, vocabularies, semantic constraints, process models, fixture
interchange artefacts, release metadata, and tests distinguish observed
evidence, candidate inference, validation results, and certified human
outcomes. This gives future FOI process research a practical base for
corpus evaluation, process analytics, release packaging, and accountable
analyst-led public-information review without claiming more than the
current repository can prove.

\section{Data and Code Availability}\label{data-and-code-availability}

The code, schemas, ontology seed, examples, documentation, and
validation contracts are maintained in the public FOI-O repository
{[}\hyperlink{ref-23}{23}{]}. Version 0.8.1 is archived in Zenodo {[}\hyperlink{ref-30}{30}{]}. The public NZ
archive dataset is distributed through the
\texttt{edithatogo/fyi-archive-nz} Hugging Face dataset repository
{[}\hyperlink{ref-24}{24}{]}, with packaging and provenance owned by \texttt{fyi-archive}
rather than FOI-O. The related programme repositories used by this work
are \texttt{fyi-cli} {[}\hyperlink{ref-4}{4}{]}, \texttt{fyi-archive} {[}\hyperlink{ref-5}{5}{]},
\texttt{foi-process} {[}\hyperlink{ref-25}{25}{]}, \texttt{nlp-policy-nz} {[}\hyperlink{ref-26}{26}{]},
\texttt{legislation} {[}\hyperlink{ref-27}{27}{]}, \texttt{rulespec-nz} {[}\hyperlink{ref-28}{28}{]}, and
\texttt{rac-conformance} {[}\hyperlink{ref-29}{29}{]}. Source request and archive content
remains subject to its original rights and platform terms.

\section{Ethics and Legal Boundary}\label{ethics-and-legal-boundary}

This work is process-support-only. It does not provide legal advice or
certify legal outcomes. Human approval is required before any
operational use that would affect FOI request handling.

\section{Author Contributions}\label{author-contributions}

Dylan A Mordaunt conceptualised the work, developed the repository,
prepared this article, and approved the current draft.

\section{Funding}\label{funding}

No specific external funding is reported in this review draft. Author
confirmation is required before submission.

\section{Conflicts of Interest}\label{conflicts-of-interest}

No conflict of interest is reported in this review draft. Author
confirmation is required before submission.

\section{References}\label{references}

\begin{enumerate}
\def\labelenumi{\arabic{enumi}.}
\tightlist
\hypertarget{ref-1}{}
\item
  Official Information Act 1982. 1982 {[}cited 2026 Jul 3{]}. Available
  from:
  \url{https://www.legislation.govt.nz/act/public/1982/0156/latest/DLM64785.html}.
\hypertarget{ref-2}{}
\item
  Office of the Ombudsman. Official Information Act guides and
  resources. 2025 {[}cited 2026 Jul 3{]}. Available from:
  \url{https://www.ombudsman.parliament.nz/resources/official-information-act-guides-and-resources}.
\hypertarget{ref-3}{}
\item
  Public Service Commission Te Kawa Mataaho. OIA statistics. 2025
  {[}cited 2026 Jul 3{]}. Available from:
  \url{https://www.publicservice.govt.nz/data/oia-statistics}.
\hypertarget{ref-4}{}
\item
  edithatogo. FYI CLI: privacy-focused CLI tool for managing FYI.org.nz
  official information requests {[}software{]}. 2026 {[}cited 2026 Jul
  3{]}. Available from: \url{https://github.com/edithatogo/fyi-cli}.
\hypertarget{ref-5}{}
\item
  edithatogo. FYI Archive {[}software{]}. 2026 {[}cited 2026 Jul 3{]}.
  Available from: \url{https://github.com/edithatogo/fyi-archive}.
\hypertarget{ref-6}{}
\item
  UNESCO. Monitoring and reporting on access to information. 2025
  {[}cited 2026 Jul 3{]}. Available from:
  \url{https://www.unesco.org/en/monitoring-access-information}.
\hypertarget{ref-7}{}
\item
  UNESCO. The Right to Information Rating. 2025 {[}cited 2026 Jul 3{]}.
  Available from:
  \url{https://www.unesco.org/en/world-media-trends/right-information-rti-rating}.
\hypertarget{ref-8}{}
\item
  Popper KR. The Open Society and Its Enemies. 2013 {[}cited 2026 Jul
  3{]}. Available from:
  \url{https://press.princeton.edu/books/paperback/9780691158136/the-open-society-and-its-enemies}.
\hypertarget{ref-9}{}
\item
  OECD. Open Government for Stronger Democracies: A Global Assessment.
  2023 {[}cited 2026 Jul 3{]}. Available from:
  \url{https://www.oecd.org/content/dam/oecd/en/publications/reports/2023/11/open-government-for-stronger-democracies_88aa0131/5478db5b-en.pdf}.
\hypertarget{ref-10}{}
\item
  OECD. Open government data: Government at a Glance 2025. 2025 {[}cited
  2026 Jul 3{]}. Available from:
  \url{https://www.oecd.org/en/publications/2025/06/government-at-a-glance-2025_70e14c6c/full-report/open-government-data_619b668c.html}.
\hypertarget{ref-11}{}
\item
  Open Government Partnership. Right to Information Performance. 2023
  {[}cited 2026 Jul 3{]}. Available from:
  \url{https://www.opengovpartnership.org/wp-content/uploads/2023/01/OGP_BL_PA_RighttoInfo_January2023.pdf}.
\hypertarget{ref-12}{}
\item
  World Bank. GovTech Maturity Index, 2022 Update: Trends in Public
  Sector Digital Transformation. 2022 {[}cited 2026 Jul 3{]}. Available
  from:
  \url{https://openknowledge.worldbank.org/entities/publication/10b535a7-e9d4-51bd-96ed-6b917d5eb09e}.
\hypertarget{ref-13}{}
\item
  OECD. OECD Survey on Drivers of Trust in Public Institutions: 2024
  Results. 2024 {[}cited 2026 Jul 3{]}. Available from:
  \url{https://www.oecd.org/en/publications/oecd-survey-on-drivers-of-trust-in-public-institutions-2024-results_9a20554b-en.html}.
\hypertarget{ref-14}{}
\item
  World Justice Project. WJP Rule of Law Index 2024: Open Government.
  2024 {[}cited 2026 Jul 3{]}. Available from:
  \url{https://worldjusticeproject.org/rule-of-law-index/global/2024/Open\%20Government/}.
\hypertarget{ref-15}{}
\item
  World Wide Web Consortium. Data Catalog Vocabulary (DCAT) - Version 3.
  2024 {[}cited 2026 Jul 3{]}. Available from:
  \url{https://www.w3.org/TR/vocab-dcat-3/}.
\hypertarget{ref-16}{}
\item
  World Wide Web Consortium. PROV-O: The PROV Ontology. 2013 {[}cited
  2026 Jul 3{]}. Available from: \url{https://www.w3.org/TR/prov-o/}.
\hypertarget{ref-17}{}
\item
  World Wide Web Consortium. Shapes Constraint Language (SHACL). 2017
  {[}cited 2026 Jul 3{]}. Available from:
  \url{https://www.w3.org/TR/shacl/}.
\hypertarget{ref-18}{}
\item
  World Wide Web Consortium. SKOS Simple Knowledge Organization System
  Reference. 2009 {[}cited 2026 Jul 3{]}. Available from:
  \url{https://www.w3.org/TR/skos-reference/}.
\hypertarget{ref-19}{}
\item
  JSON Schema. JSON Schema Draft 2020-12. 2022 {[}cited 2026 Jul 3{]}.
  Available from: \url{https://json-schema.org/draft/2020-12}.
\hypertarget{ref-20}{}
\item
  Moreau L, Groth P, Cheney J, Lebo T, Miles S. The rationale of PROV.
  Journal of Web Semantics. 2015. doi:10.1016/j.websem.2015.04.001.
\hypertarget{ref-21}{}
\item
  van der Aalst WMP, Adriansyah A, de Medeiros AKA, Arcieri F, Baier T,
  Blickle T, et al.~Process Mining Manifesto. In: Business Process
  Management Workshops. 2012. doi:10.1007/978-3-642-28108-2\_19.
\hypertarget{ref-22}{}
\item
  van der Aalst WMP. Process Mining: Data Science in Action. Springer;
  2016. doi:10.1007/978-3-662-49851-4.
\hypertarget{ref-23}{}
\item
  Mordaunt DA. FOI-O: ontology and validation stack for Freedom of
  Information process modelling {[}software{]}. 2026 {[}cited 2026 Jul
  3{]}. Available from: \url{https://github.com/edithatogo/foi-o}.
\hypertarget{ref-24}{}
\item
  Mordaunt DA. FYI Archive NZ {[}dataset{]}. Hugging Face; 2026 {[}cited
  2026 Jul 16{]}. Available from:
  \url{https://huggingface.co/datasets/edithatogo/fyi-archive-nz}.
\hypertarget{ref-25}{}
\item
  Mordaunt DA. FOI Process {[}software{]}. 2026 {[}cited 2026 Jul 16{]}.
  Available from: \url{https://github.com/edithatogo/foi-process}.
\hypertarget{ref-26}{}
\item
  Mordaunt DA. NLP Policy NZ {[}software{]}. 2026 {[}cited 2026 Jul
  16{]}. Available from:
  \url{https://github.com/edithatogo/nlp-policy-nz}.
\hypertarget{ref-27}{}
\item
  Mordaunt DA. Legislation {[}software and data{]}. 2026 {[}cited 2026
  Jul 16{]}. Available from:
  \url{https://github.com/edithatogo/legislation}.
\hypertarget{ref-28}{}
\item
  Mordaunt DA. RuleSpec NZ {[}software{]}. 2026 {[}cited 2026 Jul 16{]}.
  Available from: \url{https://github.com/edithatogo/rulespec-nz}.
\hypertarget{ref-29}{}
\item
  Mordaunt DA. RAC Conformance {[}software{]}. 2026 {[}cited 2026 Jul
  16{]}. Available from:
  \url{https://github.com/edithatogo/rac-conformance}.
\hypertarget{ref-30}{}
\item
  Mordaunt DA. FOI-O NZ: Freedom of Information ontology and process
  model. Version 0.8.1 {[}software{]}. Zenodo; 2026.
  doi:10.5281/zenodo.21360138.
\end{enumerate}

\clearpage

\section{Abbreviations}\label{abbreviations}

\begin{table}[H]
\small
\hypertarget{tab-abbreviations}{}
\begin{center}\small\textbf{Table 8: Abbreviations used in this article.}\end{center}
\begin{tabularx}{\linewidth}{>{\raggedright\arraybackslash}p{0.20\linewidth}X}
\toprule
Abbreviation & Full term \\
\midrule
BPMN & Business Process Model and Notation \\
DCAT & Data Catalog Vocabulary \\
FOI & Freedom of Information \\
FOI-O & Freedom of Information Ontology \\
FYI & FYI.org.nz public request platform \\
JSON & JavaScript Object Notation \\
JSON-LD & JavaScript Object Notation for Linked Data \\
MCP & Model Context Protocol \\
OCEL & Object-Centric Event Log \\
ODRL & Open Digital Rights Language \\
OIA & Official Information Act \\
OWL & Web Ontology Language \\
PNML & Petri Net Markup Language \\
PROV-O & Provenance Ontology \\
RDF & Resource Description Framework \\
SHACL & Shapes Constraint Language \\
SKOS & Simple Knowledge Organization System \\
XES & eXtensible Event Stream \\
\bottomrule
\end{tabularx}
\end{table}

\clearpage

\section{Glossary}\label{glossary}

\hyperlink{tab-glossary}{The glossary table} collects the key terms used
below.

\begin{table}[H]
\small
\hypertarget{tab-glossary}{}
\begin{center}\small\textbf{Table 9: Glossary of key terms used in this article.}\end{center}
\begin{tabularx}{\linewidth}{>{\raggedright\arraybackslash}p{0.28\linewidth}X}
\toprule
Term & Meaning in this article \\
\midrule
Analyst-facing & Designed so analysts can use software to read, validate, and prepare information without treating software output as a legal decision. \\
Candidate process event & A possible workflow event inferred from observed evidence and requiring review before it can be treated as certified. \\
Certified outcome & A legally meaningful outcome confirmed by an authorised human or authoritative record. \\
Controlled vocabulary & A defined list of terms used to keep states, event types, and assertions consistent. \\
External gate & A requirement that cannot be proven by the repository alone, such as live-provider verification or submission approval. \\
Certification boundary & The rule that software may support review but must not certify legal outcomes. \\
Process ontology & A machine-readable model of the steps, states, events, evidence, and review boundaries in an administrative process. \\
Provenance & Information about where a record, event, claim, or transformation came from and how it was produced. \\
Request profile & A structured description of a public-information request, including identifiers, source state, and contextual metadata. \\
Verification stack & The schemas, tests, semantic constraints, and build checks used to verify that modelled process information is well formed and bounded. \\
\bottomrule
\end{tabularx}
\end{table}

\end{document}